\documentstyle[preprint,aps,amsfonts,prd,epsf]{revtex}
\tightenlines

\begin{document}

\title {Creation of photons in an oscillating cavity with two
moving mirrors}
    
\author{Diego A.\ R.\ Dalvit  \thanks{dalvit@df.uba.ar} and
Francisco D.\ Mazzitelli \thanks{fmazzi@df.uba.ar}}

\address{ {\it
Departamento de F\'\i sica {\it J. J. Giambiagi}, 
Facultad de Ciencias Exactas y Naturales\\ 
Universidad de Buenos Aires- Ciudad Universitaria, Pabell\' on I\\ 
1428 Buenos Aires, Argentina}}

\maketitle

\begin{abstract}
We study the creation of photons in a one dimensional oscillating cavity with
two perfectly conducting moving walls. By means of a conformal transformation 
we derive a set of generalized Moore's equations whose solution contains the
whole information of the radiation field within the cavity. For the case of
resonant oscillations we solve these equations using a renormalization group
procedure that appropriately deals with the secular behaviour present in 
a naive perturbative approach. We study the time evolution of the energy 
density profile and of the number of created photons inside the cavity.

\end{abstract}

%%\pacs{12.20-m, 03.70+k, 42.50.-p}

\section{INTRODUCTION}

It is well-known that in the presence of moving boundaries the vacuum state
of the electromagnetic field may not be stable, which results in the generation
of real photons. The generated radiation exerts pressure on the moving 
boundaries which can be looked upon as a dissipative force that opposes 
itself to the mechanical motion of the boundaries.
The generation of photons, which is an amazing demonstration of the 
existence of quantum vacuum fluctuations of QED, is referred to in the 
literature as dynamical Casimir effect \cite{Schwinger} 
or motion-induced radiation \cite{LamPRL96}. 
It goes without saying that it would be very nice to have an experimental
verification of this prediction. Due to the technical difficulties involved
in the detection of the phenomenon, up to now no concrete experiment has
been carried out, and there are only some few experimental proposals
\cite{YabloPRL,DodPRA96}. However, a feasible experimental evidence is not
out of reach, and therefore it is of interest to explore different 
theoretical models to describe the process and identify signatures
which permit to distinguish vacuum radiation from spurious effects.

Research in the field has mainly concentrated on one dimensional models,
which are useful for giving an account of the main physical processes
participating in the phenomenon
(a small number of works deal with more realistic three dimensional
models \cite{DodPRA96,PauloPRA,DodPLA98}). In this work we will also restrict 
ourselves to one dimesional models. Motion induced effects of vacuum radiation
already show up for a single mirror moving with a non-uniform acceleration
in vacuum \cite{FD}. Since the amount of radiation generated is very small,
basically determined by the ratio of the speed of the mirror to the speed of
light, much attention has been paid to the study of one dimensional models 
for which the effect is enhanced.

A cavity made of two perfectly parallel reflecting mirrors, one of which is
motionless and the other one oscillates with a mechanical frequency equal
to a multiple of the fundamental optical resonance frequency of the static
cavity is a thoroughly studied example where such an enhancement takes
place  \cite{Moore,Law,Cole,MeplanPRL,DodPLA,DodJMP,SassaPRA}. 
It is typically considered that the
cavity is motionless and that at some instant one mirror starts to oscillate
resonantly with a tiny amplitude. For small times after the motion starts,
one can make a perturbative expansion of the equations of motion of the field 
in terms of the small amplitude to find an approximate solution. 
In this way one can study the structure of 
the electromagnetic field inside the cavity, which departs from
the standard static Casimir profile (which is constant over the whole 
cavity) and develops a structure of small and broad pulses. The number
of motion-induced photons grows quadratically in time, and the spectrum
has an inverted parabolic shape with an upper frequency cut-off given by the 
mechanical frequency, its maximum being at half that value 
\cite{DodPRA96,JiPRA97}. Similar results are found in \cite{LamPRL96} 
by means of a scattering approach for the radiation emitted out of a lossy 
cavity. However, for long times these methods are not valid, and new
approximation techniques are required. In \cite{DodJMP,nos}
it is shown that in such a limit the structure of the electromagnetic field
is non trivial, with a number of pulses equal to the mechanical resonant 
frequency,
whose width decreases exponentially and whose height increases exponentially
with time,
in such a way that the total energy within the cavity grows exponentially
at the expense of the energy given to the system to keep the mirror moving.
Also, the spectrum does not have an upper frequency cut-off. Through a 
process of frequency up-conversion, the generated photons contain frequencies
of higher order cavity modes and thus exceed the mechanical frequency. The
physical mechanism of such an optical pumping into the high frequency region is
the Doppler up-shift of the field upon reflection at the mirrors. Similar
conclusions are found for the lossy cavity \cite{LamQPH98}.

The case of cavitites with two moving mirrors has also been considered 
recently. In the small time approximation, both for the ideal cavity
\cite{JiPRA98} and for the lossy one \cite{LamPRL96}, it is found that the 
number of motion-induced photons grows quadratically in time and that the 
spectrum is once again parabolic.  
In the long time approximation, the lossy cavity
has been studied with the scattering approach \cite{LamQPH98}. Just as in
the case of a single oscillating mirror, it is found that in this regime,
and for the two types of motion described above, there is pulse shaping in the 
time domain and frequency up-conversion in the spectrum of emitted
photons from the cavity. A striking feature of the spectrum is that
no photons are emitted at frequencies equal to multiple integers of the
mechanical frequency.

In this work we will consider an ideal cavity with two 
mirrors oscillating resonantly at the same frequency, and we will allow for 
different amplitudes and a 
possible dephasing between the mirrors. To investigate the problem
we will deduce a generalization of Moore's equation
\cite{Moore} to the problem of two moving boundaries, whose solution
gives the complete information on the electromagnetic field inside
the cavity. For the case of resonant harmonic motions, no exact solution
exists, and we find an approximate analytic solution based on a
renormalization group (RG) technique. We have already applied this method in
\cite{nos} for the case of a single oscillating mirror, and just as in that
case, the strategy allows to find a single solution valid both for small and 
long times. This will allow us to describe precisely the bahaviour of the
energy density and the number of photons for all times.

As we shall see, motion-induced radiation strongly depends
on the relation among the amplitudes of oscillation, the frequency and
the dephasing. For some relations among these variables, there is
constructive interference and a series of pulses develop within the cavity
that grow exponentially in time, and frequency up-conversion takes place. 
For some other relations, there is destructive interference and hence no 
vacuum radiation. We also show that our solution is capable of accounting for
other physical behaviours, for which the peaks grow quadratically rather than
exponentially. 

The paper is organized as follows. In Section 2 we will introduce the 
generalization of Moore's equations for a moving cavity and we will explain 
how to calculate the energy density and the number of motion-induced photons.
In Section 3 the renormalization group method is described and applied
to the problem of harmonically oscillating walls with dephasing. 
In Section 4 we study some particular dephasings, which we will call
translational and breathing modes. In Section 5 another motion is
considered, which has a cualitatively different behaviour as compared
to those of the previous Section. Finally in Section 6 we make our
conclusions.

\section{GENERALIZED MOORE'S EQUATIONS}

We consider a one-dimensional cavity formed by two perfectly 
reflecting mirrors, each of which follows a given trajectory, say $L(t)$ for 
the left mirror and $R(t)$ for the right one. These two trajectories are
predetermined (i.e. are given data for the problem) and act as  
time-dependent boundary conditions for the electromagnetic field inside the 
cavity. The field equation for the vector potential takes the form of
the equation for a massless scalar field
$(-\partial_t^2 + \partial_x^2) A(x,t)=0$ \footnote{The speed of light is
set to unity.}
and the boundary conditions are $A(x=L(t),t)=A(x=R(t),t)=0$ for all times.
If we express the field in terms of creation $a_k^{\dagger}$ and annihilation
$a_k$ operators for photons in the form
\begin{equation}
A(x,t)= \sum_{k=1}^{\infty} \left[ a_k \psi_k(x,t) + a_k^{\dagger} 
                                   \psi_k^*(x,t) \right] ,
\end{equation}
then the mode functions $\psi_k(x,t)$ must be chosen so as to satisfy the
above boundary conditions. 

In the case when only one of the walls moves, say the right one, the modes
can be written in terms of a function $U(t)$ as 
\begin{equation}
\psi_k(x,t)=\frac{i}{\sqrt{4 \pi k}} 
\left[ e^{-i k \pi U(t+x)} - e^{-i k \pi U(t-x)} \right] ,
\end{equation}
and the boundary condition on the right is met
\footnote{The boundary condition on the left (fixed) mirror is automatically
fulfilled by this form for the modes.} provided that the function $U$
verifies $U(t+R(t))-U(t-R(t))=2$, which is known as Moore's equation
\cite{Moore}.
The complete solution to the problem involves finding a solution $U(t)$
in terms of the prescribed motion $R(t)$. Moore's equation can in fact
be deduced by means of a conformal transformation from the original spacetime
coordinates $(t,x)$ to a new set of coordinates $(\bar{t},\bar{x})$ in which
not only the left mirror but also the right one is fixed. This transformation
takes the form
\begin{eqnarray}
\bar{t} + \bar{x} = U(t+x) & ~~~ , ~~~ & \bar{t} - \bar{x} = U(t-x), 
\end{eqnarray}
which, after mapping the coordinate of the left mirror as 
$L(t)\equiv 0 \rightarrow \bar{x}_L =0$ and for the right one 
$R(t) \rightarrow \bar{x}_R =1$, leads to Moore's equation.

Let us now consider the more general case in which both mirrors move. Evidently
we can also make a similar conformal transformation, but know we need two
functions $U$ instead of one. Defining the transformation as
\begin{eqnarray}
\bar{t} + \bar{x} = G(t+x) & ~~~ , ~~~ & \bar{t} - \bar{x} = F(t-x), 
\end{eqnarray}
and mapping $L(t)$ and $R(t)$ as before, we obtain a set of generalized
Moore's equations
\begin{eqnarray}
G(t+L(t)) - F(t-L(t)) = 0 \nonumber \\
G(t+R(t)) - F(t-R(t)) = 2,  
\label{General}
\end{eqnarray}
which, when solved for given motions for the mirrors, allow us to find the
solution for the modes inside the cavity. Indeed, the modes can be casted
in the form
\begin{equation}
\psi_k(x,t)=\frac{i}{\sqrt{4 \pi k}} 
\left[ e^{-i k \pi G(t+x)} - e^{-i k \pi F(t-x)} \right] ,
\end{equation}
and they satisfy both the field equation and the boundary conditions.

We shall be interested in studying the spacetime profile of the energy
density of the field between the moving walls
\begin{equation}
\langle T_{00}(x,t) \rangle = \frac{1}{2} 
\left[
\langle \left( \frac{\partial A(x,t)}{\partial t} \right)^2 \rangle +
\langle \left( \frac{\partial A(x,t)}{\partial x} \right)^2 \rangle 
\right] ,
\end{equation}
where the expectation values are taken with respect to the vacuum state.
As is well-known, this quantity is divergent and a regularization 
method is needed to get meaningful results. Using the point-splitting
method and introducing advanced $u=t+x$ and retarded $v=t-x$ coordinates, 
the energy density can be rewritten in terms of the functions $G$ and $F$ 
in the following way \cite{FD}
\begin{eqnarray}
\langle T_{00}(u,v) \rangle &=& \frac{\pi}{4} \sum_{k=1}^{\infty}
k \left\{ 
G'(u) G'(u+i\epsilon) e^{-i k \pi [G(u)-G(u+i\epsilon)]} + \right. \nonumber \\
& & \left. ~~~~~~~~~~~~
F'(v) F'(v+i\epsilon) e^{-i k \pi [F(v)-F(v+i\epsilon)]}
\right\},
\end{eqnarray}
with $\epsilon \rightarrow 0^+$. From here it is straightforward to get 
the renormalized version, 
$\langle T_{00}(x,t) 
\rangle_{\scriptscriptstyle{\mbox{ren}}} = -f_G(t+x) - f_F(t-x)$, where

\begin{eqnarray}
f_G &=& \frac{1}{24 \pi} \left[
\frac{G'''}{G'} - \frac{3}{2} \left( \frac{G''}{G} \right)^2 +
\frac{\pi^2}{2} (G')^2  \right] \nonumber \\
f_F &=& \frac{1}{24 \pi} \left[
\frac{F'''}{F'} - \frac{3}{2} \left( \frac{F''}{F} \right)^2 +
\frac{\pi^2}{2} (F')^2  \right]. \label{T00}
\end{eqnarray}

Suppose that for $t<0$ both walls were at rest separated by a distance 
$\Lambda$ and that the field was
in its vacuum state. The solution of the generalized Moore's equations
is simply $G(t)=F(t)=t/\Lambda$ and the mode functions 
$\psi_k$ correspond to positive frequency modes. If at $t=0$ the boundaries
begin to move, it is well-known that for some types of motion the field 
does not remain in vacuum, but photons are produced through nonabiabatic 
processes. A consistent calculation of the number of created photons
through motion-induced radiation
requires to have a well-defined vacuum state in the future. To this end
we consider that at time $t=T$ both walls come to rest
($L(t)=0$ and $R(t)=\Lambda$ for $t \ge T$). The evolved old vacuum
state does not coincide with the vacuum in the future, but rather it 
contains a number of photons, which can be calculated by means of the
Bogoliubov coefficients $\beta_{nm}=(\psi^{\star}_n,\psi^{(0)}_m)^{\star}$,
where 
$\psi_m^{(0)}(x,t)=(\pi m)^{-1/2} \sin(m\pi x/\Lambda) e^{-i\pi m t/\Lambda}$ 
is the mode function for the static problem
\footnote{The inner product is the usual for Klein-Gordon equation, namely
$(\psi,\xi)=-i \int_{L(t)}^{R(t)} dx \left[ \psi \cdot{\xi}^{\star} 
- \cdot{\psi} \xi^{\star} \right] $.}, and
$\psi_n(x,t)$ is the mode function which solves the nonstationary problem 
for $t>0$ and coincides with $\psi_n^{(0)}(x,t)$ for $t<0$. Writing the
mode functions in terms of the functions $G$ and $F$, integrating by parts,
and using the set of Moore's
equations to drop the surface terms, we get the following relation for
the Bogoliubov coefficient
\begin{eqnarray}
\beta_{nm}(t,T)&=&\frac{1}{2} \sqrt{\frac{m}{n}} 
\left\{
\int_{t/\Lambda-1}^{t/\Lambda} dx \exp\{ -i\pi [n F(\Lambda x) + m x]\} +
\right. \nonumber \\
&& \left. ~~~~~~~~~~
\int_{t/\Lambda}^{t/\Lambda+1} dx \exp\{ -i\pi [n G(\Lambda x) + m x]\}
\right\},
\label{bogocoef}
\end{eqnarray}
for times $t>T$. 
The number of created photons inside the cavity after the stopping time $T$
in the $n$-th mode
is given by $N_n(T)= \sum_m |\beta_{nm}(t,T)|^2$ (the dependence of the
Bogoliubov coefficient on $t$ is just a phase) and summing over $n$ we
get the total amount of motion-induced photons.

\section{RENORMALIZATION GROUP METHOD FOR MOORE'S EQUATIONS}

For resonantly harmonic motions it is not possible to find an exact solution
to Eqs.(\ref{General}) 
and approximation methods are compelling. A naive approach
is to make perturbations in the amplitude of the oscillation, but it turns
out that the strategy is ill-fated, because of the appearance of secular terms
proportional to the time which, after a short period, make the approximation
break down. In \cite{nos} we have applied a method inspired in the 
renormalization group to treat these singular perturbations for the case
of a one dimesional cavity with one oscillating mirror. The method has a wide
range of application in different fields, especially for studying ordinary
differential equation problems involving boundary layers, multiple scales, etc.
\cite{Goldenfeld}. In the following we shall extend the method for the
case of a one dimensional cavity whose mirrors oscillate in resonance
with the cavity. More specifically, we consider that for $t<0$ the two
mirrors are motionless and separated by a distance $\Lambda$, and that
at $t=0$ they start to move as
\begin{eqnarray}
L(t) &=& \epsilon A_L \sin \left( \frac{q \pi t}{\Lambda} \right) \equiv 0 + 
\epsilon \delta L(t)
\nonumber \\
R(t) &=& \Lambda - \epsilon A_R \sin \left( \phi \right) + \epsilon A_R  
\sin \left( \frac{q \pi t}{\Lambda}+ \phi \right) \equiv 
\Lambda + \epsilon \delta R(t),
\end{eqnarray}
for the left and right mirrors respectively. Here $\phi$ is a possible 
dephasing angle, $A_L$ and $A_R$ are amplitudes of oscillation and 
$\epsilon \ll 1$ is a small parameter.

Let us first start with the perturbative approximation. We expand both  
unknown functions $G(t)$ and $F(t)$ in terms of the small parameter $\epsilon$
and retain first order terms only, $G(t)=G_0(t) + \epsilon G_1(t)$
and $F(t)=F_0(t) + \epsilon F_1(t)$. Equating terms of the same order in
the set of generalized Moore's equations, we get for the zeroth order part
\begin{eqnarray}
G_0(t) - F_0(t) &=& 0  \label{ec1}\\
G_0(t+\Lambda) - F_0(t-\Lambda) &=&2, \label{ec2} 
\end{eqnarray}
and for the first order part
\begin{eqnarray}
G_1(t) - F_1(t) &=& - \theta(t) \delta L(t) \left[ G_0'(t) + F_0'(t) \right] 
\label{ec3}\\
G_1(t+\Lambda) - F_1(t-\Lambda) &=& - \theta(t) \delta R(t) 
\left[ G_0'(t+\Lambda) + F_0'(t-\Lambda) \right]. \label{ec4}
\end{eqnarray}
The general solution to Eqs.(\ref{ec1}, \ref{ec2}) is
\begin{equation}
G_0(t)=F_0(t)= c + \frac{t}{\Lambda} + \sum_{n \ge 1} \left[
A_n \sin \left( \frac{n \pi t}{\Lambda} \right) + 
B_n \cos \left( \frac{n \pi t}{\Lambda} \right) \right],
\end{equation}
where $c$, $A_n$, and $B_n$ are constants determined by the boundary
conditions of the problem. These are obtained from the fact that the modes
$\psi_k(x,t)$ must be positive frequency modes for $t<0$, which implies
that $G(t)=t/\Lambda$ for $-\infty \le t \le \Lambda$ and
$F(t)=t/\Lambda$ for $- \infty \le t \le 0$. The different ranges for the 
functions $G$ and $F$ follow directly from the nonlocal structure of Moore's
equations. 

Making the shift $t \rightarrow t-\Lambda$ in Eq.(\ref{ec3}) and replacing
the result in Eq.(\ref{ec4}) we get an equation for the first order correction
to the function $G$, namely
\begin{eqnarray}
G_1(t+\Lambda) - G_1(t-\Lambda) &=& \theta(t-\Lambda) \delta L(t-\Lambda) 
\left[ G_0'(t-\Lambda) + F_0'(t-\Lambda) \right] - \nonumber \\
&& \theta(t) \delta R(t) \left[ G_0'(t+\Lambda) + F_0'(t-\Lambda) \right] .
\end{eqnarray}
Since this equation is linear, the solution is of the form
$G_1=G_1^{(1)}+ G_1^{(2)}$, where 
\begin{eqnarray}
G_1^{(1)}(t+\Lambda) - G_1^{(1)}(t-\Lambda) &=& \theta(t-\Lambda) 
\delta L(t-\Lambda) \left[ G_0'(t-\Lambda) + F_0'(t-\Lambda) \right] 
\label{G1} \\
G_1^{(2)}(t+\Lambda) - G_1^{(2)}(t-\Lambda) &=& 
- \theta(t) \delta R(t) \left[ G_0'(t+\Lambda) + F_0'(t-\Lambda) \right] ,
\label{G2}
\end{eqnarray}
whose general solutions read
\begin{eqnarray}
G_1^{(1)}(t) &=& \frac{A_L}{\Lambda} \frac{t}{\Lambda} \,
\theta(t) \, \sin\left( q \pi t/\Lambda \right) \times \nonumber \\
&& \left\{
 1 + \pi
\sum_{n\ge 1} n  
\left[ A_n  \cos \left(n\pi t/\Lambda \right) -
       B_n  \sin \left(n\pi t/\Lambda \right)  \right] \right\}
+  g^{(1)}(t) ,
\end{eqnarray}
for Eq.(\ref{G1}), and for Eq.(\ref{G2}) we get
\begin{eqnarray}
G_1^{(2)}(t) &=& \frac{A_R}{\Lambda} \frac{t}{\Lambda} \, \theta(t+\Lambda) \,
\left\{ \sin\left(\phi\right) + (-1)^{q+1} 
\sin\left( q \pi t/\Lambda+\phi\right) \right\} \times \nonumber \\
&& \left\{ 1 + \pi \sum_{n \ge 1} n \left[
A_n \cos \left( n \pi t/\Lambda \right) -
B_n \sin \left( n \pi t/\Lambda \right) \right] \right\} + g^{(2)}(t) ,
\end{eqnarray}
where $g^{(1)}$ and $g^{(2)}$ are arbitrary periodic functions of period 
$2 \Lambda$. The first order correction for $F$ can be deduced from the
first order correction for $G$ we have just found using Eqs.(\ref{ec3})
or (\ref{ec4}) indistinctly. We see that the perturbative corrections contain
secular terms that grow linearly in time. Therefore the approximation will be 
valid only for short times, that is, $\epsilon t/\Lambda \ll 1$.

In order to determine the two unknown periodic functions we have to 
consider the boundary conditions for the functions $G$ and $F$. 
We have already said that the nonlocal structure of Moore's equations 
implies that, although at $t=0$ the motion of the mirrors starts, 
the expression for $G$ for times up to $t=\Lambda$ is given by the solution 
for motionless walls, namely $G(t)=t/\Lambda$ for $t \le \Lambda$, while
that for $F$ reads $F(t)=t/\Lambda$ for $t \le 0$. If we assume that these 
boundary conditions are already satisfied by the zeroth-order  
solutions $G_0(t)$ and $F_0(t)$, then the periodic functions must be chosen 
so that $G_1(t)=0$ and $F_1(t)=0$ in the respective intervals. This fact,
when translated to the functions $G_1^{(1)}$ and $G_1^{(2)}$ imply the 
following boundary conditions
\begin{eqnarray}
G_1^{(1)}(t) = 0 &,& \mbox{for $0 \le t\le 2 \Lambda$} \nonumber \\
G_1^{(2)}(t) = 0 &,& \mbox{for $-\Lambda \le t \le \Lambda$} ,
\end{eqnarray}
which leads to the following expressions for the periodic functions
\begin{eqnarray}
g^{(1)}((2k+1)\Lambda+z) &=& -\frac{A_L}{\Lambda} \frac{z+\Lambda}{\Lambda}
\sin\left( q \pi z/\Lambda \right) \times \nonumber \\
&& \left\{
 1 + \pi
\sum_{n\ge 1} n  
\left[ A_n  \cos \left(n\pi z/\Lambda \right) -
       B_n  \sin \left(n\pi z/\Lambda \right)  \right] \right\},
\end{eqnarray}
and
\begin{eqnarray}
g^{(2)}(2 p \Lambda + \omega) &=& -\frac{A_R}{\Lambda} \frac{w}{\Lambda} 
\left\{ \sin\left(\phi\right) + (-1)^{q+1} 
\sin\left( q \pi \omega/\Lambda+\phi\right) \right\} \times \nonumber \\
&& \left\{ 1 + \pi \sum_{n \ge 1} n \left[
A_n \cos \left( n \pi \omega/\Lambda \right) -
B_n \sin \left( n \pi \omega/\Lambda \right) \right] \right\} ,
\end{eqnarray}
where $t=(2k+1)\Lambda+z$, $k=0,1,2,\ldots$ and $-\Lambda \le z \le \Lambda$
for the function $g^{(1)}$, while for the function $g^{(2)}$ we have
$t=2 p \Lambda+ \omega$, $p=0,1,2,\ldots$ and 
$-\Lambda \le \omega \le \Lambda$. Given $t$, the values of the integers
$k$ and $p$ are obtained as $k=p=\frac{1}{2} \mbox{int}(t/\Lambda)$ for
$\mbox{int}(t/\Lambda)$ even, and $k=\frac{1}{2} [\mbox{int}(t/\Lambda)-1]$,
$p=\frac{1}{2} [\mbox{int}(t/\Lambda)+1]$ for $\mbox{int}(t/\Lambda)$ odd.
Note that during the first period $k=0$ ($p=0$), the function $g^{(1)}$
($g^{(2)}$) makes $G_1^{(1)}$ ($G_1^{(2)}$) vanish identically. As we have 
already seen, since the mirrors were at rest for $t<0$, we must impose
$G(t)=t/\Lambda$ for $t \le \Lambda$ and $F(t)=t/\Lambda$ for $t \le 0$. 
Therefore $c=A_n=B_n=0$, and the perturbative solution for $t>0$, to order 
${\cal{O}}(\epsilon^2)$, is
\begin{eqnarray}
G(t) &=& \frac{t}{\Lambda} + \epsilon \frac{A_L}{\Lambda} 
\frac{t-z-\Lambda}{\Lambda} \sin\left( q \pi t/\Lambda \right) 
+ \epsilon \frac{A_R}{\Lambda} \frac{t-\omega}{\Lambda} 
\left[ \sin\left( \phi\right) + 
(-1)^{q+1} \sin \left( q \pi t/\Lambda + \phi \right) \right] 
\label{Gpert}\\
F(t) &=& G(t) + 2 \epsilon \frac{A_L}{\Lambda} \sin 
\left( q \pi t/\Lambda \right).
\label{Fpert}
\end{eqnarray}
These perturbative solutions suffer from the aforementioned secularity 
problems, being valid for times $t/\Lambda \ll \epsilon^{-1}$. In order to 
deal with this drawback and get improved solutions valid for longer times,
in the rest of this Section we describe the RG method we mentioned before
for this problem of an oscillating cavity.

What the renormalization group method does is to improve the perturbative 
expansion by resumming an infinite number of secular terms. In general,
if one performs the perturbative expansion to higher orders, there appear
different time scales, $\epsilon t$ to first order, $\epsilon^2 t$ and
$\epsilon^2 t^2$ to second order, $\epsilon^3 t$, $\epsilon^3 t^2$ and
$\epsilon^3 t^3$ to third order, and so on. The RG technique sums the most
secular terms of each order ($\epsilon^n t^n$), 
and it is therefore valid for times
$t/\Lambda \ll \epsilon^{-2}$. The way to carry out the resummation is
nicely described in \cite{Goldenfeld} and it basically consists in introducing
and arbitrary time $\tau$, split the time in the secular terms of the first
order perturbative corrections as $t=(t-\tau)+\tau$, and absorb the terms 
proportional to $\tau$ into the ``bare'' parameters of the zeroth order
perturbative solution, thereby becoming ``renormalized''.
Introducing the arbitrary time $\tau$ and splitting $t$ as said, the 
perturbative solution can be written as
\begin{eqnarray}
G(t,\tau)&=&c(\tau) + \sum_{n \ge 1} \left[
A_n(\tau) \sin(n \pi t/\Lambda) + B_n(\tau) \cos(n \pi t/\Lambda) \right]
+ \frac{t-\tau}{\Lambda} \nonumber \\
&&  + \epsilon \frac{t-\tau}{\Lambda}
\frac{A_L}{\Lambda} \sin\left( q \pi t/\Lambda \right) 
\left\{ 1 + \pi
\sum_{n\ge 1} n  
\left[ A_n(\tau)  \cos \left(n\pi t/\Lambda \right) -
       B_n(\tau)  \sin \left(n\pi t/\Lambda \right)  \right] \right\}
\nonumber \\
&& + \epsilon \frac{t-\tau}{\Lambda}
\frac{A_R}{\Lambda} 
\left\{ \sin\left(\phi\right) + (-1)^{q+1} 
\sin\left( q \pi t/\Lambda+\phi\right) \right\} \times \nonumber \\
&& ~~~~~~~~~~~~~~ \left\{ 1 + \pi \sum_{n \ge 1} n \left[
A_n(\tau) \cos \left( n \pi t/\Lambda \right) -
B_n(\tau) \sin \left( n \pi t/\Lambda \right) \right] \right\}
\nonumber \\
&& + \epsilon g^{(1)}(t,\tau) +\epsilon g^{(2)}(t,\tau),
\label{GRG}
\end{eqnarray}
where the bare parameters $c,A_n$ and $B_n$ have been replaced by their
renormalized counterparts $c(\tau)$, $A_n(\tau)$ and $B_n(\tau)$. Here 
$g^{(1)}(t,\tau)$ and $g^{(2)}(t,\tau)$ respectively denote the functions
$g^{(1)}(t)$ and $g^{(2)}(t)$ with the same replacement. Note that these
functions are no longer periodic due to the RG-improvement.

Since the time $\tau$ is arbitrary, the solution for $G$ should not depend
on it, which implies the following RG equation 
$(\partial G/\partial \tau)_t=0$. In our case it consists of three independent
equations
\begin{eqnarray}
\frac{\partial c(\tau)}{\partial \tau} &=& \frac{1}{\Lambda} + 
\frac{2}{\pi} (-1)^{q+1} b + {\cal O}(\epsilon^2) 
\label{difc}\\
\frac{\partial A_n(\tau)}{\partial \tau} &=&
\frac{2}{\pi} a \delta_{nq} - 2 (-1)^{q+1} b \, n B_n \nonumber \\
&& + |n-q| \left[ a A_{|n-q|} - b \, \mbox{sg(n-q)} B_{|n-q|} \right] 
\nonumber \\
&& - (n+q) \left[ a A_{n+q} + b B_{n+q} \right] + {\cal O}(\epsilon^2) 
\label{difa}\\
\frac{\partial B_n(\tau)}{\partial \tau} &=&
\frac{2}{\pi} b \delta_{nq} + 2 (-1)^{q+1} b \,n A_n \nonumber \\
&& + |n-q| \left[ a \, \mbox{sg}(n-q) B_{|n-q|} + b A_{|n-q|} \right] 
\nonumber \\
&& + (n+q) \left[ -a B_{n+q} + b A_{n+q} \right] + {\cal O}(\epsilon^2),
\label{difb}
\end{eqnarray}
where 
\begin{eqnarray}
a&\equiv& \frac{\epsilon}{\Lambda} \frac{\pi}{2} \left[
\frac{A_L}{\Lambda} + \frac{A_R}{\Lambda} (-1)^{q+1} \cos(\phi) \right] 
\label{defa}\\
b&\equiv& \frac{\epsilon}{\Lambda} \frac{\pi}{2} \frac{A_R}{\Lambda} (-1)^{q+1}
\sin(\phi). \label{defb}
\end{eqnarray}

These parameters $a$ and $b$ play a crucial role because they determine
the behaviour of the solutions to the set of generalized Moore's equations.
There are four distinct cases. The simplest one is for $a=b=0$, which happens,
for example, for equal amplitudes $A_L=A_R$, zero dephasing and even 
frequencies. In this case there is no secular behaviour at the level of
the perturbative solutions Eqs.(\ref{Gpert},\ref{Fpert}), which are then
valid also for long times. The energy inside the cavity oscillates around
the static Casimir value and there is no motion-induced radiation. 
A second case is $a\neq 0$ and $b=0$, which 
occurs, for example, for equal amplitudes, zero dephasing and odd
frequencies. In this case secular terms do appear in the perturbative 
solutions and the RG method is useful for finding the long time behaviour,
which shows an exponential increase of the energy in the cavity and 
motion-induced photons. This case will be the subject matter of the next 
Section.  A third case is $a=0$ and $b \neq 0$, which takes place, for 
example, for a static left mirror $A_L=0$ and dephasing $\phi=\pi/2$. Here 
there are also secular terms at the perturbative level, and for long times the
energy does not grow exponentially but quadratically, photons also being 
generated. We shall deal with this case in the Section V. Finally, the case 
$a \neq 0$ and $b \neq 0$ is similar to the second case, there is
motion-induced radiation and an exponential increase of the energy. We
shall not cover this case in detail since the expressions for the
solutions to Moore's equations are cumbersome.

Now we solve the RG equations (\ref{difc}-\ref{difb}).
The solution for $c$ is trivial, 
$c(\tau)=[\frac{1}{\Lambda} + \frac{\epsilon}{\Lambda} \frac{A_R}{\Lambda}
\sin(\phi) ] \tau + \kappa$, with $\kappa$ a constant to be determined.
Writing $A_n=\tilde{A}_n - \tilde{A}_{-n}$ and 
$B_n=\tilde{B}_n + \tilde{B}_{-n}$, the new variables satisfy
\begin{eqnarray}
\frac{\partial \tilde{A}_m(\tau)}{\partial \tau} &=&
\frac{2}{\pi} a \delta_{mq} - 2 (-1)^{q+1} b m \tilde{B}_m \nonumber \\
&& + (m-q) \left[ a \tilde{A}_{m-q} - b \tilde{B}_{m-q} \right] \nonumber \\
&& - (m+q) \left[ a \tilde{A}_{m+q} + b \tilde{B}_{m+q} \right] 
+ {\cal O}(\epsilon^2) \\
\frac{\partial \tilde{B}_m(\tau)}{\partial \tau} &=&
\frac{2}{\pi} b \delta_{mq} + 2 (-1)^{q+1} b m \tilde{A}_m \nonumber \\
&& + (m-q) \left[ a \tilde{B}_{m-q} + b \tilde{A}_{m-q} \right] \nonumber \\
&& + (m+q) \left[ -a \tilde{B}_{m+q} + b \tilde{A}_{m+q} \right] + 
{\cal O}(\epsilon^2). 
\end{eqnarray}

The initial conditions for these differential equations are dictated by the
perturbative solution: $c(0)=\tilde{A}_m(0)=\tilde{B}_m(0)=0$. This implies
that the constant $\kappa$ is zero. In order to solve for 
$\tilde{A}_m$ and $\tilde{B}_m$, we first decouple the equations through
the transformation $\tilde{C}_m=\tilde{A}_m - i \tilde{B}_m$ and
$\tilde{D}_m=\tilde{A}_m + i \tilde{B}_m$, and introduce a generating 
functional $M(s,\tau)=\sum_m s^m \tilde{C}_m(\tau)$. It is easy to see that
this functional verifies the following differential equation
\begin{equation}
\frac{\partial M}{\partial \tau}=\frac{2}{\pi} (a-i b) s^q +
\left[-2 i (-1)^{q+1} b s + (a-i b)s^{q+1} - (a+i b) s^{1-q}
\right] \frac{\partial M}{\partial s},
\label{Mdif}
\end{equation}
with boundary condition $M(s,\tau=0)=0$. The solution can be obtained by
proposing an Ansatz $M(s,\tau)=\Phi[ e^{-\tau} \alpha(s)] + \beta(s)$,
where $\Phi[\ldots]$, $\alpha(s)$ and $\beta(s)$ are functions to be 
determined. We shall not dwell on the details of finding these functions,
but suffice it to say that the last two are straightforwardly derived after
introducing the Ansatz in Eq.(\ref{Mdif}), while the first one is got once
the initial condition on $M$ is imposed. The solution reads
\begin{eqnarray}
M(s,\tau) &=&  \frac{2}{\pi} \left\{
i (-1)^{q+1} q b \tau - \ln \cosh(q a \tau) - \right.
\nonumber \\
&& \left. \ln \left[
\left( 1 + i (-1)^{q+1} (b/a) \tanh(q a \tau) \right) -  
\left( 1 - i b/a \right) \tanh(q a \tau) s^q \right] \right\}.
\end{eqnarray}
Expanding this solution in powers of $s$ (and doing the same for its complex
conjugate) we get to our final objective, i.e. the coefficients 
$\tilde{A}_m$ and $\tilde{B}_m$. The only non-vanishing coefficients are
\begin{eqnarray}
\tilde{A}_{m=0} &=& \frac{1}{\pi q} \left\{
-2 \ln \cosh(q a \tau) - \ln [1+i (b/a) \tanh(q a \tau)] -
\ln [1-i (b/a) \tanh(q a \tau)] \right\} \label{A0}\\
\tilde{B}_{m=0} &=& \frac{i}{\pi q} (-1)^{q+1} \left\{
2 i q b \tau - \ln [1+i (b/a) \tanh(q a \tau)] +
\ln [1-i (b/a) \tanh(q a \tau)] \right\} \label{B0} \\
\tilde{A}_{m=q j} &=& \frac{\tanh^j(q a \tau)}{\pi q j} \left\{
\frac{(1-i b/a)^j}{[1+i(-1)^{q+1} (b/a) \tanh(q a \tau) ]^j} + 
\right. \nonumber \\
&& ~~~~~~~~~~~~~~~~~~~~~~ 
\left. \frac{(1+i b/a)^j}{[1-i(-1)^{q+1} (b/a) \tanh(q a \tau) ]^j}
\right\} \label{Aqj} \\
\tilde{B}_{m=q j} &=& \frac{i \tanh^j(q a \tau)}{\pi q j} \left\{
\frac{(1-i b/a)^j}{[1+i(-1)^{q+1} (b/a) \tanh(q a \tau) ]^j} -
\right. \nonumber \\
&& ~~~~~~~~~~~~~~~~~~~~~~ 
\left.
\frac{(1+i b/a)^j}{[1-i(-1)^{q+1} (b/a) \tanh(q a \tau) ]^j}
\right\} , \label{Bqj}
\end{eqnarray}
where $j \in N$. Note that since $\tilde{A}_{m<0}=\tilde{B}_{m<0}=0$, the
original coefficients $A_m$ and $B_m$ are equal to the $\tilde{A}_m$'s and
$\tilde{B}_m$'s respectively. 

The expressions for the RG-improved coefficients ensure that the
solution for $G$ and $F$ does not depend on $\tau$. We still have the
freedom to choose the arbitrary time $\tau$ at will, and the obvious
choice is $t=\tau$, since in this way the secular terms proportional to
$t-\tau$ dissapear. Given the RG-improved coefficients, we still have to
plug them into Eq.(\ref{GRG}) and perform the necessary
summations to finally get the RG-improved solutions $G(t,t)$ and $F(t,t)$. 

For a general dephasing, the resulting expressions are rather
lengthy, so in the next two sections we will concentrate on particular cases.
Firstly we study the case of dephasing $\phi=0$, which corresponds to 
translational modes, and dephasing $\phi=\pi$, which corresponds to breathing 
modes. Secondly we analyze a case with only one 
mirror oscillating, similar to the one we studied in \cite{nos}, but with a
dephasing $\phi=\pi/2$, which gives cualitatively different results.

\section{TRANSLATIONAL AND BREATHING MODES}

In the present Section we consider that the cavity has translational
modes ($\phi=0$), or that it has breathing modes ($\phi=\pi$).
For the particular case of equal amplitudes $A_L=A_R$, the former type
of motion corresponds to the cavity oscillating as a whole, with its
mechanical length kept constant (pictorically called an ``electromagnetic 
shaker'' \cite{MeplanPhD}), while in the latter type of motion the mirrors
oscillate symmetrically with respect to the center of the cavity, the 
mechanical length changing periodically (an ``antishaker'').
Both for translational and breathing modes the expressions 
for the coefficients in Eqs.(\ref{A0}-\ref{Bqj}) 
simplify considerably because $b=0$, and the summations to get the 
functions $G$ and $F$ are straightforward. 
Setting $t=\tau$ in Eq.(\ref{GRG}) we get the RG-improved
solutions
\begin{eqnarray}
G(t,t) &=& \frac{t}{\Lambda} - \frac{2}{\pi q} \mbox{Im} 
\ln [1+ \zeta + (1-\zeta) e^{i q \pi t/\Lambda} ] + \epsilon g^{(1)}(t,t) 
+ \epsilon  g^{(2)}(t,t) 
\label{gshaker}\\
F(t,t) &=& G(t) + 2 \epsilon \frac{A_L}{\Lambda} \sin(q \pi t/\Lambda)
\frac{2 \zeta}{1+\zeta^2 + (1-\zeta^2) \cos(q \pi t/\Lambda)} ,
\label{fshaker}
\end{eqnarray}
where we have defined $\zeta \equiv \exp[2 q a t]$. The (now nonperiodic)
RG-improved functions  $g^{(1)}(t,t)$ and $g^{(2)}(t,t)$ are
\begin{eqnarray}
g^{(1)}(t,t) &=& - \frac{A_L}{\Lambda} \frac{z+\Lambda}{\Lambda}
\sin(q \pi t/\Lambda) 
\frac{2 \zeta}{1+\zeta^2 + (1-\zeta^2) \cos(q \pi t/\Lambda)} \\
g^{(2)}(t,t) &=& \mp (-1)^{q+1} \frac{A_R}{\Lambda} \frac{\omega}{\Lambda}
\sin(q \pi t/\Lambda) 
\frac{2 \zeta}{1+\zeta^2 + (1-\zeta^2) \cos(q \pi t/\Lambda)} ,
\end{eqnarray}
where, in the last formula, the upper sign corresponds to $\phi=0$ and
the lower sign to $\phi=\pi$. 
This solution for $G$ and $F$ is cualitatively similar to the one 
we obtained in the case for one oscillating wall with zero dephasing 
\cite{nos}.
For the same reasons described in that reference, both RG-improved 
nonperiodic functions give negligible corrections to $G$ and $F$ in the long
time limit ($ \epsilon^{-1} \ll t/\Lambda \ll \epsilon^{-2}$). 
However, they are crucial for the solution to satisfy the correct boundary 
conditions at short times ($t/\Lambda \ll \epsilon^{-1}$).

The energy density inside the cavity is given by Eq.(\ref{T00}) in terms of 
derivatives of $G$ and $F$. Since these expressions involve second and third
derivatives of these functions, and since there is an initial discontinuity
of the velocities of the mirrors, the energy density will develop $\delta$
function singularities that will be infinitely reflected back and forth
between the mirrors. In what follows we will ignore these singularities
\footnote{These singularities are of course artifacts of the sudden 
approximation we are using to describe the motion of the mirrors at $t=0$.}.

The structure of the electromagnetic field within the cavity  
at long times strongly depends
on the relation among amplitudes, frequencies and dephasings. 
If these are such that the coefficient $a$ is equal to zero (remember
that for the motions considered in this Section the other coefficient $b$
is always null), then there is
destructive interference. For equal amplitudes $A_L=A_R$, this takes place 
for even $q$ and dephasing $\phi=0$,
or for odd $q$ and dephasing $\phi=\pi$: all RG-improved coefficients
$A_n$ and $B_n$ are null, and there is no motion-induced radiation 
enhancement whatsoever. 
If, on the contrary, $a \neq 0$, then we have constructive interference,
which, for $A_L=A_R$, is maximal for odd $q$ and $\phi=0$, or for
even $q$ and $\phi=0$. Radiation enhancement takes place: the electromagnetic 
shaker and antishaker have ``explosive cocktails`` at long times. 
In particular, for $q \ge2$, the RG solutions $G(t,t)$ and $F(t,t)$ develop
a staircase form. Within regions of $t$ between odd multiples of $\Lambda$,
there are a total of $q$ jumps located at values of $t$ 
for which the argument of the logarithm in Eq.(\ref{gshaker}) vanishes,
i.e.
$\cos(q \pi t/\Lambda) = \pm 1$, where the upper sign corresponds to $a>0$
and the lower one to $a<0$. 
In Figure 1 we show the form of the functions $G$ and $F$ for
short times and in Figure 2 for long times. 
Note that in the long time limit they are practically the same. 
The energy density builds up
a number of $q$ travelling wave packets which become narrower as
$\exp(-2 q |a| t)$ and higher as $\exp(4 q |a| t)$, so that
the total energy inside the cavity grows like 
$\exp(2 q |a| t)$ at the expense of the energy pumped into the system to keep
the mirrors moving as predetermined. 
In Figure 3 the profile of the energy density inside the cavity at a fixed 
time is depicted. We compare the case of the shaker with that of a single 
oscillating mirror. The difference in height and width of the peaks between
these two situations are due to the fact that the parameter $a$ for the shaker
is twice that of the single mirror. This reflects how the cavity can enhance
vacuum radiation. 

A rather different picture appears when one considers the $q=1$ case, 
which corresponds to an oscillation frequency
equal to the lowest eigenfrecuency of the cavity. In this case the energy does
not grow exponentially, but oscillates around the static Casimir value. 

Now we turn to calculate the number of motion-induced  photons inside
the cavity. We assume that before the mirrors started to move the state of the
field was vacuum, and that at time $t=T$, when both walls come once again to 
rest
\footnote{For the motions $L(t)$ and $R(t)$ we are considering in this Section,
this happens for times $T$ such that $T/\Lambda=2 k/q$, ($k \in N$).},
we define a new vacuum, in which there is an amount of real photons given by 
the Bogoliubov coefficients Eq.(\ref{bogocoef}). 
To calculate these coefficients at a
time $t>T$ we need to know the form for the function $G$ and $F$ in the 
corresponding time intervals as appear in the integral expression 
Eq.(\ref{bogocoef}).

To this end let us discuss briefly how the RG-improved solutions $G(t,t)$
and $F(t,t)$ match the solutions to the problem of motionless walls for
$t>T$. The non-local structure of Moore's equation implies that the solution
for $F(t)$ is the RG one $F(t,t)$ up to $t\le T$, and that for $G(t)$ is the 
RG one $G(t,t)$ up to $t \le T+\Lambda$. Also, evaluating Moore equation for
times $t \ge T$ it follows that $F(t)=G(t)$ for $t \ge T$. Finally, for
$t \ge T+\Lambda$ both Moore's equations can be combined to obtain the usual
equation for a static cavity, so $G(t)=F(t)=t/\Lambda + \Delta(t)$, where
$\Delta(t)$ is a $2 \Lambda$-periodic function that we must determine. If
due care is taken of the boundary conditions at the moment when the walls stop,
it is easy to see that this function can be written by periodizing the
RG-improved functions $G(t,t)$ and $F(t,t)$ as follows
\begin{equation}
\Delta(t=T+2\Lambda+z) = \left\{
\begin{array}{ll}
F(T+z,T+z) & \mbox{for} -\Lambda \le z \le 0 \\
G(T+z,T+z) & \mbox{for} 0 \le z \le \Lambda
\end{array}
\right. ,
\end{equation}
and $\Delta(t)=\Delta(t+2 \Lambda)$.

Having now the form of the solutions for times after the stopping of the
walls, we can calculate the Bogoliubov coefficients for late times
$t/\Lambda >T/\Lambda \gg \epsilon^{-1} $ 
in a manner similar to that of \cite{DodJMP}. 
Let us split the solution $G(t)$ in Eq.(\ref{gshaker}) 
in the form $G(t)=G_s(t) + G_{\mbox{np}}(t)$, where the first part 
is $G_s(t)=t/\Lambda - 2/(\pi q) \mbox{Im} 
\ln [1+ \zeta + (1-\zeta) e^{i q \pi t/\Lambda} ]$ and the last part involves
the RG-improved functions  $g^{(1)}$ and $g^{(2)}$. 
From a graph of these nonperiodic functions
and of the difference between $G(t)$ and $F(t)$ one can see that, for long
times, they are all bounded, much smaller than $G_s$, and that they are
effectively zero except in small time intervals which tend to zero as time
increases. Therefore we can drop their contribution in the imaginary 
exponents of the integral representation of the Bogoliubov coefficient,
and get
\begin{equation}
\beta_{nm}(t,T)=\frac{1}{2} \sqrt{\frac{m}{n}} 
\int_{t/\Lambda-1}^{t/\Lambda+1} dx \exp\{ -i\pi [n G_s(\Lambda x) + m x]\}.
\label{bog}
\end{equation}
The function $G_s$ has a first term, linear in time, and a second one, that
for late times becomes an oscillating function, its period being 
$2\Lambda/q$ and the amplitude of its oscillations being independent 
of $\epsilon$. Then the Bogoliubov coefficient can be rewritten as follows
\footnote{As we have anticipated, the dependence of the Bogoliubov coefficient
on the time $t>T$ is just a phase.}
\begin{equation}
\beta_{nm}(t,T) = \frac{1}{2} \sqrt{\frac{m}{n}} e^{-i \pi (n+m) (t/\Lambda-1)}
\sum_{k=0}^{q-1} e^{-i \pi (n+m) (2/q) k} 
\int_0^{2/q} dx e^{-i \pi [ (n+m) x + n f(x) ]},
\label{bogper}
\end{equation}
with 
$f(x)=-2/(q \pi) \mbox{Im} \ln[1+\zeta + (1-\zeta) \exp(i q \pi x)]$,
and $\zeta=\exp(2 q a T)$.
To go further and be able to perform the integral we make a piecewiese 
linear approximation for the function $f$, valid for late times.
We concentrate on the case $a<0$, for which $\zeta \rightarrow 0$ 
at late times \footnote{The case $a>0$ gives similar results for the amount
of created particles. The technical difference is that since 
$\zeta \rightarrow \infty$ for late times, the approximate function is 
different.}.
From the graph of $f(x)$ one can see that it can be approximated
by
\begin{equation}
\tilde{f}(x) = \left\{ \begin{array}{ll}
- (1-q \delta) x & \mbox{for} 0 \le x \le \frac{1}{q} - \delta \\
-\frac{1}{q \delta} (1-q \delta)^2  \left( x-\frac{1}{q} \right) & \mbox{for}
\frac{1}{q} - \delta \le x \le \frac{1}{q} + \delta \\
- (1-q \delta) \left( x - \frac{2}{q} \right) & 
\mbox{for}
\frac{1}{q} + \delta \le x \le \frac{2}{q} 
\end{array}
\right. ,
\end{equation}
where $\delta=2 \sqrt{\zeta}/(q \pi)$. 
With this approximate form the integrals become trivial, and  
after neglecting the integral over the middle interval which is proportional 
to $\delta$, one can get a closed expression for the Bogoliubov coefficient,
valid for $m \delta \ll 1$. For the particular case $q=2$ we get
\begin{equation}
|\beta_{nm}(T)|^2 = \frac{m}{n \pi^2} [1+(-1)^{m+n}]
\frac{1-(-1)^m \cos(2 \pi n \delta)}{(m+2 \pi n \delta)^2}.
\end{equation}
Next we need to calculate the sum over $n$ in order to find the amount
of motion-induced photons in the $n$-th mode after the stopping time $T$. 
Using the summation formulas of \cite{DodJMP} we get
\begin{equation}
N_m(T)= \frac{1}{m^2} \left[ \ln \left( \frac{m}{2 \delta}\right)
- (-1)^m \ln \left( \frac{1}{2 \pi \delta} \right) \right].
\end{equation}
Recalling that $\delta$ is a function of $T$, and taking the $T$-derivative
we find the rate of photoproduction
\begin{equation}
\frac{d N_m(T)}{d T}= - \frac{2 a}{m \pi^2} [1-(-1)^m].
\end{equation}
These results are valid for $\epsilon T/\Lambda \gg 1$ and not for very large
wave numbers $m \delta \ll 1$
\footnote{There is a further restriction that comes from the fact that we
are using a sudden approximation for the motion of the mirrors at $t=0$ and
$t=T$. Indeed, if we assume that $\tau_s$ is the characteristic time for the
mirror to come to rest, the sudden approximation will be valid for modes such
that $m \ll \Lambda/\tau_s$.}.
The number of photons per mode grows linearly in the stopping time and the 
rate, for late times, approaches an asymptotic value that depends on the 
value of $a$, i.e. one the relation among amplitudes, frequency and dephasing. 
Both for the shaker and the antishaker, motion-induced radiation is
enhanced in comparison to the case of a single oscillating mirror in a
cavity. Indeed, in the former cases the rate of photoproduction is twice
that of the latter case. Photons are created in the odd modes only, 
whereas their amount in even mode is zero (it may be different from zero
in the next-to-leading order approximation). 
This situation is typical of processes involving
parametric excitations \cite{DodJMP}.

\section{A DEPHASED OSCILLATING BOUNDARY}

In this Section we discuss another particular motion of the walls, namely one 
for which the left mirror is static and the right one oscillates resonantly,
with a dephasing $\phi=\pi/2$. The motion we consider in then
$R(t)=\Lambda - 2 \epsilon A_R \sin^2(q \pi t/2 \Lambda)$, which for
$q=2$ corresponds to the small $\epsilon$-expansion of an exact solution
to Moore's equation studied in \cite{Law}.
Our motivation for studying this peculiar case is that for this motion
we have $a=0$ and $b \neq 0$, which, as we have anticipated, gives
cualitatively different physical results.

The expressions for the functions $G(t,t)$ and $F(t,t)$ are obtained taking the
limit $a \rightarrow 0$ of Eqs.(\ref{A0}-\ref{Bqj}). The result is
\begin{eqnarray}
G(t,t)&=&F(t,t) = \frac{t}{\Lambda_{\scriptsize \mbox{eff}}} - 
\frac{2}{\pi q} {\mbox {Im}} \ln \left[
1- \frac{i q b t  e^{i q \pi t/\Lambda}}{1-i (-1)^{q+1} q b t}  \right]
+ \epsilon g^{(2)}(t,t) 
\label{Glaw}\\
g^{(1)}(t,t) &=& 0 \\
g^{(2)}(t,t) &=& - \frac{A_R}{\Lambda} 
\frac{1+(-1)^{q+1} \cos(q \pi t/\Lambda)}{1+2 q b t \sin(q \pi t/\Lambda) 
+ 2 (q b t)^2 [1+(-1)^{q+1} \cos(q \pi t/\Lambda)]} .
\end{eqnarray}
where $\Lambda_{\scriptsize \mbox{eff}}=\Lambda (1-\epsilon A_R/\Lambda)$ 
is the time-averaged length of the cavity for $t>0$.

The solution $G(t,t)$ develops a staircase
profile, the jumps being located at values of $t$ for which the argument of
the logarithm in Eq.(\ref{Glaw}) vanishes, i.e. for 
$\cos(q \pi t/\Lambda)=\pm 1$, where the plus sign corresponds to even $q$
and the minus sign to odd $q$. The energy density for this type of motions 
also consists of a series of $q$ peaks that travel between the mirrors. The
cualitative difference is that in this case the height of the peaks grows
as $(q b t)^4$, their width decreases as $(q b t)^{-2}$ and the total energy 
contained in the cavity grows quadratically rather than exponentially. 
This follows from the fact that time enters in the logarithm of Eq.(\ref{Glaw})
as a power law instead of an exponential, as in 
Eqs.(\ref{gshaker},\ref{fshaker}).

Next we calculate the amount of motion-induced radiation for 
this case. To this end we assume that at time $t=T$ the wall comes to rest,
$R(t)=\Lambda_{\scriptsize \mbox{eff}}$ for $t \ge T$, where $T$ 
is of the form $T/\Lambda=(2k+1)/(2q), (k\in N)$. This choice for the stopping
time simplifies the computation of the the Bogoliubov coefficients 
$\beta_{nm}$. In such a case Eq.(\ref{bog}) is slightly modified
\begin{equation}
\beta_{nm}(t,T)=\frac{1}{2} \sqrt{\frac{m}{n}} 
\int_{t/\Lambda_{\scriptsize 
\mbox{eff}}-1}^{t/\Lambda_{\scriptsize \mbox{eff}}+1} dx 
\exp\{ -i\pi [n G_s(\Lambda_{\scriptsize \mbox{eff}}x) + m x]\}.
\end{equation}
Now $G_s$ consists of the first two terms of Eq.(\ref{Glaw}), the first being
linear in time and the second one being an oscillating function for late
times, whose period is $2 \Lambda/q$. The Bogoliubov coefficient can also be
rewritten in a way similar to Eq.(\ref{bogper})
\begin{equation}
\beta_{nm}(t,T) = \frac{1}{2} \sqrt{\frac{m}{n}} 
e^{-i \pi (n+m) (t/\Lambda_{\scriptsize \mbox{eff}}-1)}
\sum_{k=0}^{q-1} e^{-i \pi (n+m) (2/q) k} 
\int_0^{2/q} dx \, e^{-i \pi [ (n+m) x + n f(x) ]},
\label{betalaw}
\end{equation}
since in the interval $[t/\Lambda_{\scriptsize 
\mbox{eff}}-1, t/\Lambda_{\scriptsize \mbox{eff}}+1]$ there are a total
of $q (\Lambda_{\scriptsize \mbox{eff}}/\Lambda) \approx q$ periods.
Here
\begin{equation}
f(x)=- \frac{2}{\pi q}
{\mbox {Im}} \ln \left[ 1 - 
\frac{i q b \Lambda_{\scriptsize \mbox{eff}} x 
e^{i q \pi x \Lambda_{\scriptsize \mbox{eff}}/\Lambda}}
{1-i (-1)^{q+1} q b \Lambda_{\scriptsize \mbox{eff}} x} \right].
\end{equation}
The piecewiese linear approximation for the function $f$ is in this case
\footnote{We concentrate on even frequencies, for which $b<0$. For odd 
frequencies, the results are similar.}
\begin{equation}
\tilde{f}(x) = \left\{ \begin{array}{ll}
\left( \frac{2}{q \delta} - \frac{5}{2} \right) x - \frac{1}{q}
& \mbox{for} 0 \le x \le \delta \\
- \left(1 - \frac{3}{4} q \delta \right) \left(x -\frac{2}{q} \right) -
\frac{1}{q}
& \mbox{for}
\delta \le x \le \frac{2}{q}
\end{array}
\right. ,
\end{equation}
where $\delta =[q \pi (q b T)^2]^{-1} \ll 1$ for late times. Now the integral
in Eq.(\ref{betalaw}) is straightforward, 
and after dropping the integral over the
first interval $[0,\delta]$, which is proportional to $\delta$, one can
get a closed expression for the Bogoliubov coefficients, valid as long as
$m \delta \ll 1$. For comparison with \cite{Law} we concentrate on the case
$q=2$. In this case we have
\begin{equation}
|\beta_{nm}(T)|^2 = \frac{2m}{n \pi^2} [1+(-1)^{m+n}]
\frac{1-(-1)^m \cos(3 \pi n \delta/2)}{(m+3\pi n \delta/2)^2} .
\end{equation}
Finally, we perform the summation over $n$ to get the number of created 
photons in the $n$-th mode. Using the same summation formulas as in 
\cite{DodJMP}
we get
\begin{equation}
N_m(T)=\frac{2}{m \pi^2} \left[ 
\ln \left( \frac{2m}{3\delta} \right) - (-1)^{m}
\ln \left( \frac{2}{3 \pi \delta} \right) \right].
\end{equation}
Replacing the value for $\delta$ and taking the $T$-derivative, we get
the following formula for the rate of photon production, valid in the limits
$m \delta \ll 1$ and $\epsilon T/\Lambda \gg 1$
\begin{equation}
\frac{d N_m(T)}{dT} = \frac{4}{m \pi^2} [1-(-1)^m] \frac{1}{T}.
\end{equation}
We see that the number of photon per mode grows logarithmically in 
the stopping time, and
in consequence the rate of photon creation decreases towards zero.
Similarly to the case of the vibrating cavity, photons are produced only in
odd modes.

\section{CONCLUSIONS}

In this paper we have presented a unified and analytic treatment of the
dynamical Casimir effect in a one dimensional resonantly oscillating cavity 
for arbitrary amplitudes and dephasings. We have derived a generalization
of Moore's equation to describe the state of the electromagnetic field 
inside the cavity with two moving mirrors. 
Using a technique inspired in the renormalization group 
method, we have found a solution to the set of generalized Moore's equations
which is valid both for short and long times. The physical behaviour of the
moving cavity depends crucially on the relation among amplitudes, frequency
and dephasing. We have shown that for certain cases there is destructive
interference and no radiation is generated. For others, there is constructive
interference and motion-induced photons appear. When this takes place,
the way the energy within the cavity and the number of created photons grow
in time depends on the relation among the above variables. For
certain motions the growth of the energy density is exponentially and for 
some others it is a power law.

We hope in the future to apply the RG method to more realistic 
situations, such as three dimensional oscillating cavities with rectangular 
or spherical shapes.

\section{Acknowledgments}

D. D. would like to thank Luis E. Oxman and Paulo A. Maia Neto 
for enlightning discussions on
related matters, and the hospitality of the Centro Brasileiro de Pesquisas
Fisicas, where part of this work was carried out. 
This research was supported by Universidad de Buenos Aires,
Consejo Nacional de Investigaciones Cient\'\i ficas y T\' ecnicas
and by Fundaci\' on Antorchas.

{\bf Note added:} While we were finishing this manuscript we received a paper
\cite{DodQPH} where the problem of photon creation in a cavity with two
moving mirrors is analyzed using a different method. It is shown that if the
frequency of the vibrations is not exactly a resonant one, photoproduction
is highly supressed for strong detuning.

%%%%%%%%%%%%%%%%%%%%%%%%%%%%%%%%%%%%%%%%%%%%%%%%%%%%%%%%%%%%%%%%%%%%%%%%%%%%%

%%%%%%%%%%%%%%%%%%%%%%%%%%%%%%%%%%%%%%%%%%%%%%%%%%%%%%%%%%%%%%%%%%%%%%%%%%%

%FIGURE CAPTIONS

\begin{figure}[h]
\centering \leavevmode
\epsfxsize=10cm
\epsfbox{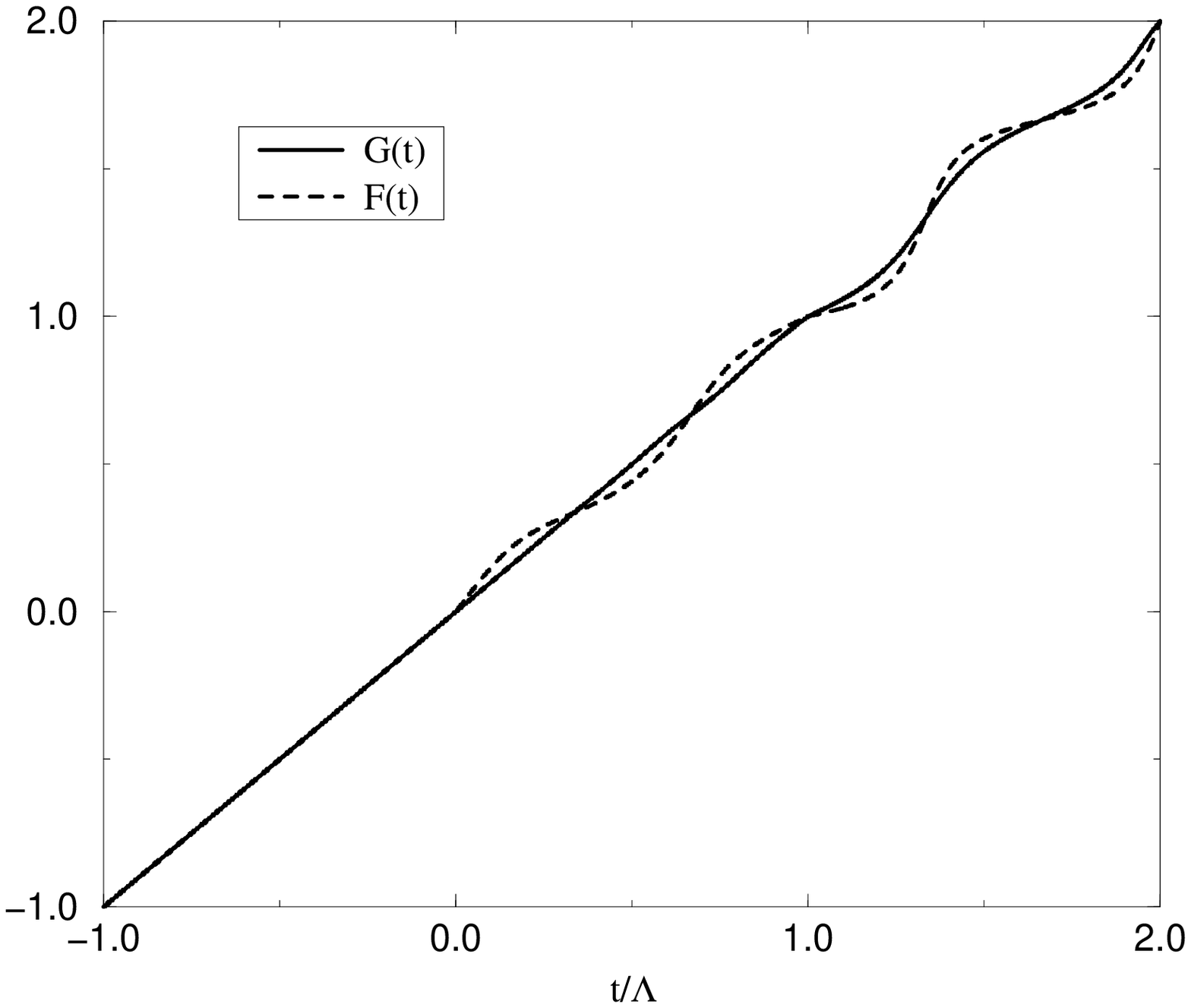}
\caption{\small{$G(t)$ and $F(t)$ vs. $t/\Lambda$ as given by 
Eqs.(\ref{gshaker},\ref{fshaker}) for small times $q |a| t/\Lambda \ll 1$.
Note that the function $F(t)$ departs from the straight line at $t=0$, while
the function $G(t)$ departs from it at $t=\Lambda$, as dictated by the initial
boundary conditions. The parameters are those for a shaker with
$A_L/\Lambda=A_R/\Lambda=1$, $q=3$, $\phi=0$ and $\epsilon=0.03$.}}
\end{figure}

\newpage

\begin{figure}[h]
\centering \leavevmode
\epsfxsize=10cm
\epsfbox{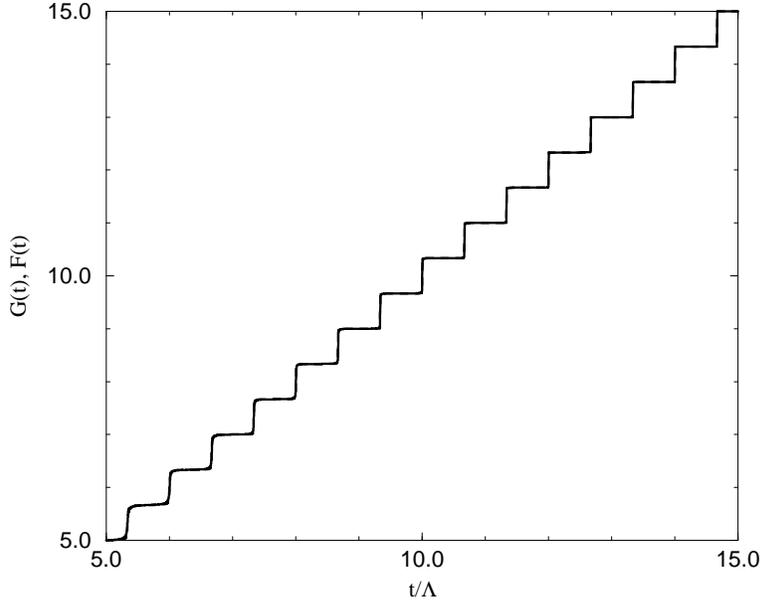}
\caption{\small{$G(t)$ and $F(t)$ vs. $t/\Lambda$ as given by 
Eqs.(\ref{gshaker},\ref{fshaker}) for long times $q |a| t/\Lambda \gg 1$.
At these times both functions coincide and take a staircase profile. 
The parameters are the same as in figure 1.}}
\end{figure}

\begin{figure}[h]
\centering \leavevmode
\epsfxsize=10cm
\epsfbox{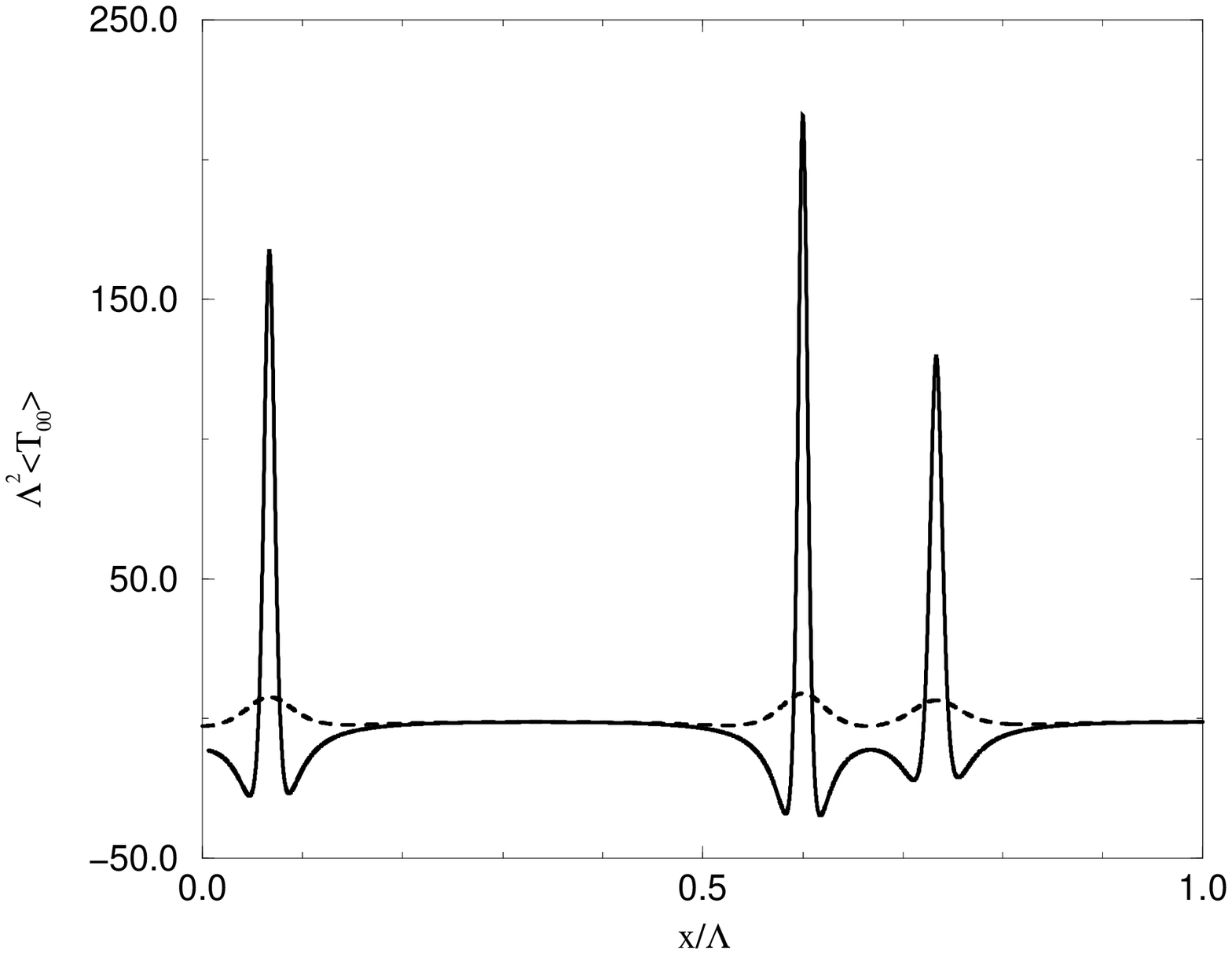}
\caption{\small{Energy density profile between plates for fixed time 
$t/\Lambda=15.4$. The solid line corresponds to a shaker with 
$A_L/\Lambda=A_R/\Lambda=1$, $q=3$, $\phi=0$ and $\epsilon=0.01$. The 
dashed line corresponds to a cavity with a single oscillating mirror ($A_L=0$) 
with the same parameters. Note that the height of the peaks for the shaker
grow as $\exp(4\pi q \epsilon t/\Lambda)$, while for the single mirror
as $\exp(2\pi q \epsilon t/\Lambda)$.}}
\end{figure}


\begin{references}

\bibitem{Schwinger} J. Schwinger, Proc. Natl. Acad. Sci. USA 
{\bf 90}, 958 (1993); {\it ibid} {\bf 90} 2105 (1993); 
{\it ibid} {\bf 90} 4505 (1993); {\it ibid} {\bf 90} 7285 (1993).

\bibitem{LamPRL96} A. Lambrecht, M.-T Jaekel and S. Reynaud, Phys. Rev. Lett.
{\bf 77}, 615 (1996).

\bibitem{YabloPRL} E. Yablonovitch, Phys. Rev. Lett. {\bf 62}, 1742 (1989).

\bibitem{DodPRA96} V. V. Dodonov and A. B. Klimov, Phys.\ Rev.\ A {\bf 53}, 
2664 (1996).

\bibitem{PauloPRA} D. F. Mundarain and P. A. Maia Neto, Phys. Rev. A {\bf 57}
1 (1998).

\bibitem{DodPLA98} V. V. Dodonov, Phys. Lett. A {\bf 244} 517 (1998).

\bibitem{FD} S. A. Fulling and C. W. Davies, Proc.\ R.\ Soc.\ Lond.\ A 
{\bf 348}, 393 (1976).

\bibitem{Moore} G. T. Moore, J.\ Math.\ Phys.\ {\bf 11}, 2679 (1970).

\bibitem{Law} C. K. Law, Phys.\ Rev.\ Lett.\ {\bf 73}, 1931 (1994).

\bibitem{Cole} C. K. Cole and W. C. Schieve, Phys. Rev. A {\bf 52}, 
4405 (1995).

\bibitem{MeplanPRL} O. M\'eplan and C. Gignoux, Phys. Rev. Lett. {\bf 76} 
408 (1996).

\bibitem{DodPLA} V. V. Dodonov, A. B. Klimov and V. I. Man'ko, 
Phys.\ Lett.\ A {\bf 149},
225 (1990).

\bibitem{DodJMP} V. V. Dodonov, A. B. Klimov and D. E. Nikonov,
J.\  Math.\ Phys.\ {\bf 34}, 
2742 (1993).

\bibitem{SassaPRA} E. Sassaroli, Y. N. Srivastava and A. Widom,
Phys. Rev. A {\bf 50} 1027 (1994).

\bibitem{JiPRA97} J.-Y. Ji, H.-H. Jung, J.-W. Park and K.-S. Soh, 
Phys. Rev. A {\bf 56} 4440 (1997).

\bibitem{nos} D. A. R. Dalvit and F. D. Mazzitelli, Phys. Rev. A {\bf 57}, 
2113 (1998).

\bibitem{LamQPH98}  A. Lambrecht, M.-T Jaekel and S. Reynaud, 
quant-ph/9805044.  

\bibitem{JiPRA98} J.-Y. Ji, H.-H. Jung and K.-S. Soh, 
Phys. Rev. A {\bf 57} 4952 (1998).

\bibitem{Goldenfeld} L. Y. Chen, N. Goldenfeld and Y. Oono, 
Phys.\ Rev.\ Lett.\ {\bf 73},
1311 (1994); Phys.\ Rev.\ E {\bf 54}, 376 (1996). 

\bibitem{MeplanPhD} O.  M\'eplan, PhD Thesis, {\it Ondes et particules dans
le modele de l'accelerateur de Fermi. Simulation numerique}, Universite
Joseph Fourier-Grenoble I, France (1996).

\bibitem{DodQPH} V. V. Dodonov, quant-ph/9810077.

\end{references}
\end{document}